\documentclass[twocolumn,superscriptaddress,amsfont,amssymb,amsmath, showpacs,balancelastpage, nofootinbib]{revtex4-1}
\usepackage{graphicx,longtable,mathrsfs,color,array}
\usepackage[hidelinks]{hyperref}
\usepackage[usenames,dvipsnames]{xcolor}
\usepackage{amssymb,amsmath,mathtools,mathrsfs}
\usepackage{epsfig,subfigure,placeins,float}
\usepackage{booktabs,longtable,ctable,multirow}
\usepackage{exscale,relsize}
\usepackage[normalem]{ulem}
\usepackage{enumerate}
\usepackage{times, mathptmx}
\usepackage{colortbl} 
\usepackage[dvipsnames]{xcolor} 
\usepackage{pifont} 
\usepackage{amsmath}

\newcommand{\be}{\begin{equation}}
\newcommand{\ee}{\end{equation}}

\newcommand{\Gtwo}{G_2{}}
\newcommand{\Gthree}{G_3{}}
\newcommand{\Gfour}{G_4{}}
\newcommand{\Gfive}{G_5{}}

\newcommand{\Ffour}{F_4{}}
\newcommand{\Ffive}{F_5{}}

\newcommand\alphaH{\alpha_{\text{H}}}

\newcommand\f{F}
\newcommand\fX{{F_X}}
\newcommand\p{P}
\newcommand\q{Q}
\newcommand\A{A}
\newcommand\B{B}
\newcommand\h{A_3}
\newcommand\Am{{\cal A}}
\newcommand\M{{\cal M}}
\newcommand\GN{G_{\rm N}}

\allowdisplaybreaks

\begin{document}
 
\title{{Scalar-tensor theories and modified gravity  in the wake of GW170817}}
\author{David Langlois}
\affiliation{APC (CNRS-Universit\'e Paris-Diderot), 10 rue Alice Domon et L\'eonie Duquet, 75205 Paris, France}
\author{Ryo Saito}
\affiliation{Graduate School of Science and Engineering, Yamaguchi University, Yamaguchi 753-8512, Japan}
\author{Daisuke Yamauchi}
\affiliation{Faculty of Engineering, Kanagawa University, Kanagawa, 221-8686, Japan}
\author{Karim Noui}
\affiliation{Laboratoire de Math\'ematiques et Physique Th\'eorique (UMR CNRS 7350), Universit\'e Fran\c cois Rabelais, Parc de Grandmont, 37200 Tours, France}
\affiliation{APC (CNRS-Universit\'e Paris-Diderot), 10 rue Alice Domon et L\'eonie Duquet, 75205 Paris, France}

 \date{\today}

\begin{abstract}
Theories of dark energy and modified gravity can be strongly constrained by astrophysical or cosmological observations, as  illustrated by the recent observation of the gravitational wave event GW170817 and of  its electromagnetic counterpart GRB 170817A, which shows that the speed of gravitational waves, $c_g$, is the same as  the speed of light, within deviations of order $10^{-15}$.
This observation implies severe restrictions on scalar-tensor theories, in particular theories whose action depends on second derivatives of a scalar field. Working in the very general framework of Degenerate Higher-Order Scalar-Tensor (DHOST) theories, which encompass Horndeski and Beyond Horndeski theories, we present the DHOST theories that satisfy $c_g=c$. We then examine, for these theories, the screening mechanism that suppresses scalar interactions on small scales, namely the Vainshtein mechanism, and  compute  the corresponding gravitational laws  for a non-relativistic spherical body. We show that it can lead to a deviation from standard gravity inside matter, parametrized by three coefficients  which satisfy a consistency relation and  can be constrained by present and future astrophysical observations. 

\end{abstract}

\maketitle

\vskip.1cm
\emph{Introduction.}  
The combination of the gravitational wave event GW170817 \cite{TheLIGOScientific:2017qsa}, observed by the LIGO/Virgo collaboration, and of  the gamma-ray burst GRB 170817A   \cite{Monitor:2017mdv} gives a remarkably precise measurement of the speed of gravitational waves (GWs): it coincides with the speed of light with deviations smaller than a few parts in $10^{-15}$.  This measurement has dramatic consequences on the viability of theories of    dark energy and/or modified gravity that have been intensively studied in the last few years because many of them generically predict a speed of GW that differs from the speed of light. 

The  purpose of this work is to reassess the viability of existing scalar-tensor theories with this new information, in the context of the most general framework that has been developed so far to describe scalar-tensor theories, namely that of Degenerate Higher-Order Scalar-Tensor (DHOST) theories, introduced in  \cite{Langlois:2015cwa} (see also \cite{Langlois:2017mdk} for a brief overview), which include and extend previously known families of theories such as Horndeski~\cite{Horndeski:1974wa} and Beyond Horndeski~
\cite{Gleyzes:2014dya}. 

Since the speed of gravitational waves for all DHOST theories has already been computed, it is a rather simple task to infer the consequences of the constraint $c_g=c$ on DHOST theories, and, consequently, on Horndeski and Beyond Horndeski. In what follows, we  present a summary of the present status of scalar-tensor theories in the wake of GW170817. We then turn to the more difficult task of deriving, for these DHOST theories satisfying $c_g=c$, the equations governing the gravitational behaviour inside and around a non-relativistic spherical object. Whereas standard gravity is recovered outside the object, we find that the gravitational equations {\it inside} matter exhibit deviations from standard gravity. These deviations  can be confronted to astrophysical 
observations in order to further constrain this large family of scalar-tensor theories.

\vskip.1cm
\emph{Overview of DHOST theories.}  
Let us  first present DHOST theories, which at present constitute  the most general class of scalar-tensor theories that propagate a single scalar degree of freedom. DHOST theories were introduced  in \cite{Langlois:2015cwa}, where it was understood  that the presence of a single scalar degree of freedom in higher-order scalar-tensor theories is the consequence of a special degeneracy of the Lagrangian, which prevents the appearance of a (problematic) extra scalar degree of freedom, even if the Euler-Lagrange equations are not necessarily second-order. All DHOST theories with a quadratic dependence on second derivatives were obtained in \cite{Langlois:2015cwa} by using this degeneracy condition. Their action can be written in the form 
\be
\label{action}
\begin{split}
S &= \int d^4 x \sqrt{-g} \left[\f(\phi, X) \, {}^{(4)}\!R+\p(\phi,X)+\q(\phi,X)\Box \phi\;, 
\right.
\\
& \left. \qquad\qquad\qquad 
+\sum_{I=1}^5\A_I(\phi, X) \,  L_I\right]\;,
\end{split}
\ee
where we have introduced, in the second line, all possible  Lagrangians that are quadratic in second derivatives of $\phi$, i.e. in  $\phi_{\mu\nu}\equiv \nabla_{\nu}\nabla_\mu \phi$ (combined with the metric $g_{\mu\nu}$ or the scalar field gradient  $\phi_\mu\equiv \nabla_\mu \phi$): 
\be
\begin{split}
& L_1 = \phi_{\mu \nu} \phi^{\mu \nu} \,, \quad
L_2 =(\Box \phi)^2 \,, \quad
L_3 = (\Box \phi) \phi^{\mu} \phi_{\mu \nu} \phi^{\nu} \,,  \\
& L_4 =\phi^{\mu} \phi_{\mu \rho} \phi^{\rho \nu} \phi_{\nu} \,, \quad
L_5= (\phi^{\mu} \phi_{\mu \nu} \phi^{\nu})^2\,,
\end{split}
\ee
each multiplied by a function $\A_I$ of $\phi$ and $X\equiv g^{\mu\nu}\nabla_\mu \phi \nabla_\nu \phi$. In order to get a theory propagating a single degree of freedom, i.e. a DHOST theory,  the functions $\f$ and $\A_I$ are restricted by degeneracy conditions, explicitly given in \cite{Langlois:2015cwa}. One can further enlarge the space of theories by adding to the above action (\ref{action}), terms of the form~\cite{BenAchour:2016fzp}
\be
\label{L_cubic}
L_{\rm cubic}=f_{(3)} (\phi,X) \, G_{\mu\nu} \phi^{\mu\nu}+\sum_{I=1}^{10} \B_I(X,\phi)\,  L^{(3)}_I\,,
\ee
where the $L^{(3)}_I$ are the ten possible combinations with a cubic dependence on $\phi_{\mu\nu}$ (we will not need their explicit form here). By imposing the degeneracy criterium, all DHOST theories combining quadratic and cubic terms have been systematically classified in \cite{BenAchour:2016fzp}. 

Among the DHOST theories thus identified, one can recognize  as particular cases the Beyond  Horndeski theories. 
Following the notation of \cite{Langlois:2015cwa}, 
they are characterized by 
\be
\label{bH}
\f=\Gfour\,, \  \A_1=-\A_2=2 \Gfour_X+ X \Ffour\,, \  \A_3=-\A_4=2 \Ffour\,,\ 
\A_5=0\,,
\ee
and $f_{(3)}=\Gfive$ with the $\B_I$ related to $\Gfive_X$ and $\Ffive$,  while $\p$ and $\q$ coincide with $\Gtwo$ and $\Gthree$, respectively. The Beyond Horndeski subclass itself  contains the Horndeski theories, obtained by simply taking $\Ffour=\Ffive=0$.
 Another subclass of DHOST theories was found earlier in \cite{Zumalacarregui:2013pma}, via disformal transformations of the Einstein-Hilbert action. 

Since DHOST theories constitute the  most general class of scalar-tensor theories considered so far, and  encompass all previously studied scalar-tensor theories as well as new ones, they provide a natural arena in which to study the viability of scalar-tensor theories. A first step in this direction was performed recently by studying scalar and tensor linear perturbations around a homogeneous and isotropic background~\cite{Langlois:2017mxy}. It was found that, for most classes of DHOST theories classified in \cite{BenAchour:2016fzp}, the square of the propagation speed of tensor modes   and that of the scalar mode have opposite sign, which implies  that a gradient instability develops in either the scalar or tensor sector. Only one class of DHOST theories, dubbed class I, which includes Horndeski and beyond Horndeski, does not suffer from this problem. This class contains four independent functions of $\phi$ and $X$ (not including the functions $\p$ and $\q$).  The theories of class I can in fact be related to Horndeski theories via disformal transformations~\cite{BenAchour:2016fzp}, but as soon as matter is taken into account, it is much simpler to use the DHOST formulation where matter is {\it minimally} coupled, as we will be briefly comment in the conclusion. 

\vskip.1cm
\emph{DHOST theories satisfying $c_g=c$.}  
Let us now turn to the new constraint brought by the discovery of a GW event and its optical counterpart, showing that the speed of gravitational waves and that of light coincide to a very high precision. For simplicity here, we will ignore the possibility of some (extreme) fine-tuning and assume that this observation is a strong indication    that $c_g$ and $c$  {\it strictly} coincide. 

Since the speed of gravitational waves (with respect to a cosmological background) has already been computed for all DHOST theories in \cite{Langlois:2017mxy}, it is a straightforward exercise to identify all  DHOST theories that ``survive'' after GW170817. In fact, it is even more useful to consider the ADM form of the DHOST Lagrangian, also derived in  \cite{Langlois:2017mxy} (in a gauge where the scalar field is spatially uniform), because it provides nonlinear information and thus makes the  background-dependent effects manifest. Concentrating on the  terms
in the Lagrangian that contribute to the kinetic and spatial gradient terms  of the tensor modes, one finds~\cite{Langlois:2017mxy}
\be
\label{L_gw}
L_{\rm ADM}= (\f - X\A_1)\, K_{ij}K^{ij}+\f \,   {}^{(3)}\!R+\dots
\ee
The requirement $c_g=c$ for any background thus imposes the very simple condition 
\be
\A_1=0\,.
\ee 
The function $\A_3$ is arbitrary, like $F$,  and the remaining functions $\A_2$, $\A_4$ and $\A_5$ are then fixed by the three degeneracy conditions derived   in  \cite{Langlois:2015cwa}, so that we have 
\be
\label{A_cg=1}
\begin{split}
& \A_1 = \A_2=0\,,  \\
& \A_4=\frac{1}{8\f}\left[ 48 \fX^2 -8(\f-X\fX) \A_3-X^2 \A_3^2\right] \,, \\
& \A_5 =\frac{1}{2 \f}\left(4\fX+X \A_3\right) \A_3\,.
\end{split}
\ee
Moreover, some of the cubic terms of (\ref{L_cubic}) also contribute to the gravitational speed,  in  a background-dependent way, and in order to eliminate this contribution, all cubic terms  must be discarded, as a consequence of the degeneracy conditions for cubic terms obtained in \cite{BenAchour:2016fzp}. 

In summary, the condition $c_g=c$ dramatically restricts the set of viable DHOST theories but still allows a   total  Lagrangian that depends on four arbitrary functions of $X$ and $\phi$, namely $\f$,   $\p$, $\q$ and $\A_3$,  given by
\be
\label{DHOST}
\begin{split}
& L^{_{\rm DHOST}}_{c_g=1}=  \p + \q\,  \Box\phi +  \f  \, {}^{(4)}\! R +  \h\phi^\mu \phi^\nu \phi_{\mu \nu} \Box \phi  \\
&+\frac{1}{8\f} \bigg(48 \fX^2 -8(\f-X\fX) \h-X^2 \h^2 \bigg) \phi^\mu \phi_{\mu \nu} \phi_\lambda \phi^{\lambda \nu} \\
&+\frac{1}{2 \f}\left(4\fX+X \h\right) \h(\phi_\mu \phi^{\mu \nu } \phi_\nu)^2 \;.
\end{split}
\ee

It is then straightforward to infer how the condition $c_g=c$ restricts the beyond Horndeski and Horndeski subclasses. Indeed, the condition $\A_1=0$ combined  with (\ref{bH}) implies $\Ffour=-2 \Gfour_X/X$, 
while the cubic terms vanish, i.e. $\Ffive=\Gfive=0$. This corresponds to the above Lagrangian (\ref{DHOST}) with  $\A_3=-4\fX/X$.
As for Horndeski theories (with $\Ffour=0$), they are restricted to $\Gfour=\Gfour(\phi)$. 
Note that the consequences of GW170817 on scalar-tensor theories were also discussed in \cite{Creminelli:2017sry,Ezquiaga:2017ekz,Baker:2017hug}, but from a narrower perspective focussed on Horndeski or Beyond Horndeski theories, which does not allow, in the most general case, for a direct inclusion of matter in the physical frame (i.e. with minimal coupling to the metric).  Here, the matter Lagrangian can be added immediately to our Lagrangian (\ref{DHOST}).

\vskip.1cm
\emph{Vainshtein mechanism in DHOST theories.} 
The  main purpose of the present work is to focus on another crucial  aspect that viable theories must satisfy, namely that standard gravity should be approximately recovered in the situations where gravity has been well tested. 
In this respect, it is essential that the scalar interactions that arise in a scalar-tensor theory are suppressed on small scales in order to get  a gravitational behaviour in agreement with solar system tests. 

For theories with higher order derivatives of the scalar field, it is well known that this is achieved by the so-called Vainshtein mechanism \cite{Vainshtein:1972sx} (see e.g. \cite{Babichev:2013usa} for a review). The Vainshtein mechanism has been investigated in detail in Horndeski theories~\cite{Kimura:2011dc,Narikawa:2013pjr,Koyama:2013paa}. For Beyond Horndeski theories, \cite{Kobayashi:2014ida} found the surprising result that 
the Vainshtein mechanism works only partially in the sense that the usual gravitational law is modified {\it inside matter}. This specific deviation from standard gravity was  studied for  various astrophysical  objects in order to obtain constraints on the parameters of the Beyond Horndeski models  (see e.g. \cite{Koyama:2015oma, Saito:2015fza,Sakstein:2015zoa, Sakstein:2015aac,Jain:2015edg,Sakstein:2016ggl,Babichev:2016jom,Sakstein:2016oel}), as recently summarized in \cite{Sakstein:2017xjx}. 
So far however, the Vainshtein mechanism has not been studied for  the most general 
DHOST theories, and  the main goal of this work is to remedy this situation. 

In order to study the Vainshtein mechanism, we consider a nonrelativistic spherical object, characterized by the mass density $\rho(r)$. The presence of this object will induce a (small) deformation of spacetime, which is described by the perturbed  metric
\be
ds^2=-(1+2\Phi(r))dt^2+(1-2\Psi(r))\delta_{ij}dx^i dx^j\,,
\ee
where we have introduced the two gravitational potentials $\Phi(r)$ and $\Psi(r)$. Our goal is to determine the gravitational field, i.e. $\Phi$ and $\Psi$, generated by the mass distribution $\rho(r)$. 
The scalar field $\phi$ also acquires a perturbation due to the presence of the object and can be written as 
\be
\phi=\phi_c(t)+\chi(r)\,,
\ee
where $\phi_c(t)$ denotes the cosmological scalar field. 
Here, in contrast to the metric, we need to include the cosmological dependence of the scalar field because it is relevant even for a local observer, as we will see below. However, we still neglect the time dependence of $\chi$, as we are interested in spatial scales much smaller than the Hubble radius, i.e. $Hr\ll1$.

The relations between the two metric perturbations, the scalar perturbation and the mass distribution can be obtained by writing down the equations of motion for the scalar field and for the metric derived from the action (\ref{action}),
following the same procedure as that used in \cite{Kobayashi:2014ida}.
These equations are expanded up to linear order in the metric perturbations, which  are small for a nonrelativistic object. We also assume $\chi^{\prime 2}\ll \dot\phi_c^2$,  but we keep terms with higher-order derivatives of $\chi$  since they become important in the Vainshtein mechanism.
Using the notation $v \equiv \dot{\phi}_c$, $x\equiv \chi'/r$, $y\equiv \Phi'/r$, $z\equiv \Psi'/r$, and 
\be
\Am (r)\equiv\frac{\M (r)}{8\pi r^3}\,, \qquad \M (r)\equiv 4\pi\int_0^r\bar r^2\rho(\bar r){\rm d}\bar r\,,
\ee
this approximation leads to the following three equations, obtained,  after integration over $r$, respectively from the scalar field equation of motion and  the time and radial components  of the metric equations of motion:
\be
\label{EOM}
\begin{split}
& (8\A_3+6\A_4)x^3+2 (\A_3+\A_4)r x\left(6x x'+rx^{\prime 2}+r xx''\right)
\\
&\  \ +\left(12\fX-(\A_3-6\A_4)v^2\right) x\, y+\left(4\fX+(\A_3+2\A_4)v^2\right)r x\, y'
\\
&\ \ -24\fX\, x \, z - 8\fX r x\,  z'=0
  \,,  \\
& \left(4\fX+(3\A_3+2\A_4)v^2\right)x^2+\left(4\fX+(\A_3+2\A_4)v^2\right)r x x'
\\
&\ \ + 2v^2(4\fX+\A_4 v^2)y-4(\f+2\fX v^2)z+2\Am=0\,,
\\
& 2\fX(x^2+rxx') +(\f+2\fX v^2)y- \f z=0\,,
\end{split}
\ee
where we have furthermore neglected terms with lower powers of $x$, since we are interested in regions inside or around the central object, thus deep inside the Vainshtein radius where $x$ is large.
The terms $\A_I$, $\f$ and $\fX$ appearing in (\ref{EOM}) are all evaluated on the background.

One can then use the last two  equations of (\ref{EOM}) to express $y$ and $z$ in terms of $x$ and $\Am$. Substituting  these expressions into the first one leads to  an equation involving only the functions $x$ and $\Am$. This equation a priori involves derivatives of $x$ but, remarkably, substituting the expression for $\A_4$ given in (\ref{A_cg=1}) yields an equation involving no derivative of $x$, which reads
\be
\begin{split}
&x\, \Big\{4 \f \A_3 (4\f+4\fX v^2-3\A_3 v^4)x^2
\\
&\quad -\left[3(4\fX+\A_3 v^2)^2v^2+4\f(12\fX+7\A_3v^2)\right]\Am
\\
&\quad -(4\fX+\A_3 v^2)(4\f+4\fX v^2+\A_3 v^4)r\Am'\Big\}=0\,.
\end{split}
\ee
Concentrating on the nontrivial solution, the above expression yields $x^2$ in terms of the matter function $\Am$ and its derivative $\Am'$. Substituting back into the earlier expressions for $y$ and $z$, it is then easy to write $y$ and $z$ in terms of $\Am$ and its derivatives. This gives the modified gravitational laws 
\be
\label{gravity}
\begin{split}
\frac{d\Phi}{dr}&= \frac{\GN \M(r)}{r^2}+\Xi_1 \GN \M''(r)\,,
\\
\frac{d\Psi}{dr}&= \frac{\GN \M(r)}{r^2}+\Xi_2\frac{\GN \M'(r)}{r}+\Xi_3\,  \GN \M''(r)\,,
\end{split}
\ee
with the effective Newton's constant  $\GN$  defined by the expression
\be
(8\pi\GN)^{-1}={2\f+2\fX v^2- \frac32\A_3 v^4}\equiv  2\f (1+\Xi_0)
\ee
 and the dimensionless coefficients
\be
\label{Xi}
\Xi_1=-\frac{(4\fX+\A_3 v^2)^2}{16\f \A_3}\,,\
\Xi_2=\frac{2\fX v^2}{\f}\,, \ \Xi_3=\frac{16\fX^2-\A_3^2 v^4}{16\A_3 \f}\,.
\ee

Outside the matter source, $\M$ is constant and one recovers the usual  gravitational behaviour. But, inside the matter distribution,  the above equations differ from those of standard gravity  in a way that is reminiscent of Beyond Horndeski theories~\cite{Kobayashi:2014ida}. Indeed,  the equation for $\Phi$ is very similar to that of Beyond Horndeski since it exhibits an additional dependence on $\M''(r)$ but the relation between the coefficient $\Xi_1$ and the functions that appear in the initial Lagrangian is different. The contrast with Beyond Horndeski is even more pronounced in the equation for $\Psi$ since we see a  new term, which depends on the second derivative of $\M$: such a term has never been considered before\footnote{
In the Beyond Horndeski case, we have $4\fX+X \A_3=0$, as discussed below (\ref{DHOST}). As a consequence, 
$\Xi_3=0$ and  $\Xi_2=-\Xi_0=-2\Xi_1$.}. 
In this respect,  the gravitational laws (\ref{gravity}) significantly extend those discussed in [27], which are restricted to Beyond Horndeski models.

Note that both $\Xi_1$ and $\Xi_3$ vanish in the special case where $4\fX- X\A_3=0$, but the deviation from standard gravity is still present via $\Xi_2$. To cancel simultaneously all $\Xi_i$ requires going to the trivial case where all $\A_I$ vanish and $\f=\f(\phi)$. 

Interestingly, the three coefficients in (\ref{Xi}) are not independent but satisfy the consistency relation
\be
\label{consistency}
\Xi_3^2-\Xi_1^2=\frac12 \Xi_1\, \Xi_2\,.
\ee
It is also worth noting that
the coefficients $\Xi_i$ can be related to two of the  cosmological effective parameters, namely  $\alphaH$ and $\beta_1$, introduced in  the effective description of dark energy in   \cite{Gleyzes:2014rba} and 
\cite{Langlois:2017mxy}, extending  the parameters of \cite{Bellini:2014fua} to  Beyond Horndeski and DHOST theories.
One finds the very simple relations: 
$\Xi_0=-\alphaH-3 \beta_1$,  and 
\be
\Xi_1=-\frac{(\alphaH+\beta_1)^2}{2(\alphaH+2\beta_1)}\,,\  \Xi_2=\alphaH\,, \  \Xi_3=-\frac{\beta_1(\alphaH+\beta_1)}{2(\alphaH+2\beta_1)}\,.
\ee

For physical situations where only the metric potential $\Phi$ is relevant,  one can re-use for $\Xi_1$ the constraints already obtained  in previous studies devoted to Beyond Horndeski theories\footnote{Our parameter $\Xi_1$ numerically coincides with $\Upsilon_1/4$, where $\Upsilon_1$ is the parameter used  in several  works based on Beyond Horndeski. We prefer to use a different notation here to emphasize that the underlying theories are different and to avoid possible confusions in the interpretation of data.}.
 In particular, the relation for $\Phi$ leads to a modified density profile for non-relativistic stars \cite{Koyama:2015oma, Saito:2015fza}. As shown in \cite{Saito:2015fza}, one must always have  $\Xi_1> -1/6$ to guarantee the very existence of  stars. One can also get the  upper bound  
 $\Xi_1\lesssim 0.4$  
 from the  consistency of the minimum mass for hydrogen burning in stars with the lowest mass red dwarf~\cite{Sakstein:2015zoa,Sakstein:2015aac}. 
 
 The ratio between the Newton's constant $\GN$ appearing in (\ref{gravity}) and the effective Newton's constant associated with the propagation of gravitational waves, given by   $8\pi G_{\rm gw}\equiv 1/(2F)$ [as can be read from (\ref{L_gw}) with $\A_1=0$] coincides with $(1+\Xi_0)^{-1}$. As discussed in \cite{Jimenez:2015bwa}, this ratio can be strongly constrained by the decay of the orbital period of the Hulse-Taylor binary pulsar since the energy loss via gravitational radiation is proportional to $G_{\rm gw}$. The present agreement with the GR predictions yields a tight constraint on $\Xi_0$, estimated  at below the percent level in \cite{Jimenez:2015bwa,Dima:2017pwp}, although a more precise analysis would be useful.  Note that a  constraint on $\Xi_0$  much tighter than on the other $\Xi_i$ effectively leads to a one-parameter approximation of the deviations from standard gravity, with $\Xi_0\approx 0$, $\Xi_1\approx 2\beta_1$, $\Xi_2\approx -3\beta_1$ and $\Xi_3\approx -\beta_1$ (and $\alphaH\approx -3\beta_1$), where, according to the constraints on $\Xi_1$ discussed earlier,  $\beta_1$ is restricted to the range $-1/12 < \beta_1\lesssim 0.2$.

Additional constraints on the $\Xi_i$ can be obtained via gravitational lensing observations, which are directly sensitive to the combination $\Phi+\Psi$, as discussed in \cite{Narikawa:2013pjr,Koyama:2015oma,Sakstein:2016ggl} in the restricted case of Horndeski and Beyond Horndeski theories. For instance, lensing in  galaxy clusters, assuming a Navarro-Frenk-White profile~\cite{Navarro:1996gj} $\rho(r)= \rho_s/[\frac{r}{r_s} (1+ \frac{r}{r_s})^2]$, is governed by 
 \begin{eqnarray}
 \Phi+\Psi&=& \frac{8\pi \GN \rho_sr_s^3}{r}\left[- \ln\left(1+\frac{r}{r_s}\right) +(\Xi_1+\Xi_3)\frac{r^2}{2(r+r_s)^2} \right.
 \cr
 && \left.\qquad\qquad \qquad \qquad  -\Xi_2\frac{r}{2(r+r_s)}\right]\,,
 \end{eqnarray}
 where the two additional terms, arising from modified gravity, behave very differently from the standard term and, moreover, have a  distinct radial dependence at small radii $r<r_s$. 
The analysis of existing and future gravitational lensing data could thus provide  interesting complementary constraints on the parameters of viable DHOST theories.

\medskip 

\emph{Conclusions.} 
The most general scalar-tensor theory propagating a single scalar degree of freedom and compatible with the observation of GW170817 is described by the DHOST action (\ref{DHOST}), which depends on four arbitrary functions of $X$ and $\phi$ (including the functions $P$ and $Q$, which do not affect the speed of gravitational waves). 

We have derived the modified gravitational laws due to the Vainshtein mechanism  for  all 
 DHOST theories verifying $c_g=c$. We have found new consequences of the breaking of the Vainshtein mechanism inside matter, leading to modified gravitational laws  that are 
 more general than those associated with the restricted case of  Beyond Horndeski theories. 
  It is crucial to use these new extended results to confront comprehensively   modified gravity models 
 with astrophysical observations depending on both gravitational potentials.

Note that our analysis has been carried out in the physical frame (or Jordan frame) where matter is minimally coupled to the metric. As shown in {\cite{Crisostomi:2016czh,Achour:2016rkg}, the quadratic DHOST theories of class I, thus including (\ref{DHOST}), can be reformulated as Horndeski theories via a disformal transformation of the metric, of the form
${\tilde g}_{\mu\nu}=C(X, \phi) g_{\mu\nu}+D(X, \phi) \, \phi_\mu\, \phi_\nu$,
 but at the price of introducing a disformal coupling of  matter to the metric. Although a study of the Vainshtein mechanism in such a ``Horndeski frame'' is feasible, it turns out that it is significantly more involved than the present analysis in the physical frame~\cite{LSY17}. 

It will be interesting in the future to obtain  quantitative constraints on the parameters $\Xi_i$, via astrophysical and cosmological observations, keeping in mind the consistency relation (\ref{consistency}). Another important question is the status of compact objects in these modified gravity models. Neutron stars in Beyond Horndeski have been studied in \cite{Babichev:2016jom,Sakstein:2016oel} but the  particular 
models considered in those works do not satisfy  $c_g=c$ and it would thus be important to reexamine these studies in the context of the viable DHOST theories discussed here.

\vskip.1cm
\emph{Acknowledgements:} 
This work was supported in part by Grant-in-Aid from
the Ministry of Education, Culture, Sports, Science and Technology (MEXT) of Japan,  No.~17K14304 (D.~Y) and No.~17K14286. (R.~S).


 \bibliographystyle{utphys}
\bibliography{MG_DHOST_biblio_new}

\end{document}